\documentclass[10pt,conference]{IEEEtran}

\usepackage{hyperref}
\usepackage{amsmath,amsthm,amssymb,amsfonts}
\usepackage{mathtools}
\usepackage{bm}
\usepackage{algorithmic}
\usepackage{graphicx}
\usepackage{textcomp}
\usepackage{xcolor}
\usepackage{braket}
\usepackage[inline]{enumitem}
\def\BibTeX{{\rm B\kern-.05em{\sc i\kern-.025em b}\kern-.08em
    T\kern-.1667em\lower.7ex\hbox{E}\kern-.125emX}}
\usepackage[natbib=true, style=ieee, sorting=none]{biblatex}
\usepackage[capitalize]{cleveref}
\usepackage{subfigure}
\usepackage{booktabs}
\usepackage{fancyhdr}

\newtheorem{theorem}{Theorem}[section]

\newtheorem{definition}[theorem]{Definition}

\newcommand{\vect}[1]{\mathbf{#1}}

\newcommand{\normal}[0]{\mathcal{N}}
\newcommand{\norm}[1]{\vert \vert {#1} \vert \vert}

\counterwithin*{theorem}{section}
\newtheorem{innercustomgeneric}{\customgenericname}
\providecommand{\customgenericname}{}
\newcommand{\newcustomtheorem}[2]{%
  \newenvironment{#1}[1]
  {%
   \renewcommand\customgenericname{#2}%
   \renewcommand\theinnercustomgeneric{##1}%
   \innercustomgeneric
  }
  {\endinnercustomgeneric}
}

\newcustomtheorem{customthm}{Theorem}
\newcustomtheorem{customlemma}{Lemma}
\newcustomtheorem{customproposition}{Proposition}

\def\eqref#1{equation~\ref{#1}}

\def\ceil#1{\lceil #1 \rceil}

\def\1{\bm{1}}

\DeclareMathAlphabet{\mathsfit}{\encodingdefault}{\sfdefault}{m}{sl}
\SetMathAlphabet{\mathsfit}{bold}{\encodingdefault}{\sfdefault}{bx}{n}

\newcommand{\E}{\mathbb{E}}

\DeclareMathOperator{\argmax}{\rm{argmax}}

\DeclarePairedDelimiterX{\infdivx}[2]{[}{]}{%
  #1\;\delimsize\|\;#2%
}

\begin{document}

\title{Quadratic Advantage with Quantum Randomized Smoothing Applied to Time-Series Analysis}

\fancypagestyle{specialfooter}{%
  \fancyhf{}
  \renewcommand\headrulewidth{0pt}
  \fancyfoot[R]{ \noindent\fbox{%
    \parbox{\textwidth}{%
        {\footnotesize \copyright 2024 IEEE. Personal use of this material is permitted. Permission from IEEE must be obtained for all other uses, in any current or future media, including reprinting/republishing this material for advertising or promotional purposes, creating new collective works, for resale or redistribution to servers or lists, or reuse of any copyrighted component of this work in other works.}
        }
    }}
}

\author{
    \IEEEauthorblockN{Nicola Franco\IEEEauthorrefmark{2}, Marie Kempkes\IEEEauthorrefmark{3}, Jakob Spiegelberg\IEEEauthorrefmark{3}, Jeanette Miriam Lorenz\IEEEauthorrefmark{2}}
    \IEEEauthorblockA{\IEEEauthorrefmark{2}Fraunhofer Institute for Cognitive Systems IKS, Munich, Germany
    \\\{nicola.franco, jeanette.miriam.lorenz\}@iks.fraunhofer.de}
     \IEEEauthorblockA{\IEEEauthorrefmark{3}Volkswagen Group Innovation, Volkswagen AG, Wolfsburg, Germany
    \\\{marie.kempkes, jakob.spiegelberg\}@volkswagen.de}
}

\maketitle
\thispagestyle{plain}
\pagestyle{plain}
\thispagestyle{specialfooter}

\begin{abstract}
As quantum machine learning continues to develop at a rapid pace, the importance of ensuring the robustness and efficiency of quantum algorithms cannot be overstated. Our research presents an analysis of quantum randomized smoothing, how data encoding and perturbation modeling approaches can be matched to achieve meaningful robustness certificates. By utilizing an innovative approach integrating Grover's algorithm, a quadratic sampling advantage over classical randomized smoothing is achieved. This strategy necessitates a basis state encoding, thus restricting the space of meaningful perturbations. We show how constrained $k$-distant Hamming weight perturbations are a suitable noise distribution here, and elucidate how they can be constructed on a quantum computer. The efficacy of the proposed framework is demonstrated on a time series classification task employing a Bag-of-Words pre-processing solution. The advantage of quadratic sample reduction is recovered especially in the regime with large number of samples. This may allow quantum computers to efficiently scale randomized smoothing to more complex tasks beyond the reach of classical methods.
\end{abstract}


\section{Introduction}

The integration of quantum computing (QC) principles into machine learning, forming the field of quantum machine learning (QML), represents a groundbreaking shift in the landscape of computational intelligence~\cite{wittek2014quantum, schuld2015introduction, biamonte2017quantum}.
Theoretically, QML offers remarkable potential benefits, including the ability to handle exponentially larger datasets and to perform certain computations much faster than classical algorithms~\cite{abbas2021power, cerezo2022challenges}.
However, QML is not without its vulnerabilities, particularly in the realm of adversarial attacks~\cite{lu2020quantum, west2023benchmarking, franco2024predominant}.
These are instances where input data is deliberately manipulated to deceive the model into making incorrect predictions or classifications, a challenge well-documented in the classical machine learning domain~\cite{biggio2013evasion, szegedy2014intriguing}.
In response to these adversarial threats, both quantum and classical machine learning communities have developed probabilistic~\cite{cohen2019certified, salman2019provably, weber2021, huang2023, sahdev2023} and deterministic~\cite{guan2021robustness, guan2022verifying, franco2022quantum, franco2023efficient} defense strategies.
\smallskip

Defenses are typically developed as either formal verification of outcomes via input space propagation or by utilizing sampling distributions linked to differential privacy. 
While the latter method has shown promise for enhancing QML applications, it is not without its limitations. 
For instance, the assumptions made by Weber~\textit{et al.}~\cite{weber2021} about class probability estimates overlook the complexities of machine learning uncertainties, rendering their robustness certificates potentially overly optimistic. 
Moreover, their claims of robustness face challenges due to issues with data encoding and scalability, raising doubts about their usefulness in practical scenarios.
Similarly, Du \textit{et al.}'s~\cite{du2021} work is limited by its focus on depolarization noise, thereby restricting the range of applicable data encodings.
Their approach does not accurately represent realistic perturbation distributions, particularly for basis state encodings, leading to a mismatch between theoretical robustness and practical applicability. 
Huang~\textit{et al.}~\cite{huang2023} while considering rotation noise, also suffers from similar shortcomings, with robustness certifications heavily dependent on data encoding types and potentially inadequate for large-scale quantum systems.
A more detailed discussion of these works can be found below in section \ref{sec:qml_robustness}.
\smallskip

These critiques highlight a gap between the theoretical robustness measures proposed and their practical utility across various QML applications. 
The failure to integrate perturbation types with data encodings may hinder the achievement of realistic robustness certifications. 
Our method aligns with QuAdRo~\cite{sahdev2023} due to its flexibility regarding the algorithm and noise type, and its clear demonstration of how QC reduces the sample size needed for robustness certification. 
However, it also has limitations: it requires data to be encoded in basis states due to the use of the Grover oracle, limiting data encoding options.
On this background, we investigate how data encoding and perturbation types have to be matched to get meaningful, and with QuAdRo scalable, robustness certificates.
Our contributions can be summarized as follows:

\begin{itemize}
    \item The connection between information encoding and perturbation type is discussed at length, we show that the interpretation of certificates obtained with randomized smoothing changes significantly and might even be meaningless for some combinations.
    \item Constrained $k$-distant Hamming weight perturbations are identified as meaningful perturbation type compatible with a basis state encoding as necessary for QuAdRo. Additionally, an algorithm for implementing them on a quantum computer is introduced.
    \item We demonstrate quadratic sampling advantage using QuAdRo on a practical time series classification task by utilizing an established Bag-of-Words pre-processing technique.
\end{itemize}
\section{Preliminaries}

\subsection{QC background}

QC operates quite differently from traditional computing. At its core, QC uses the quantum bit, or qubit, symbolized as $\ket{\psi}$. 
What makes a qubit unique is its capacity to simultaneously represent two states, $\ket{0}$ and $\ket{1}$, thanks to a property called superposition, which is described by the equation $\ket{\psi} = \psi_1 \ket{0} + \psi_2 \ket{1}$. 
Superposition allows a qubit to exist in a complex space, the Hilbert space, with dimensions that grow exponentially as $2^n$, where $n$ is the total number of qubits.

In contrast to classical bits that represent a single state at a time, qubits' ability to embody dual states simultaneously allows quantum computers to process multiple computations in parallel. 
This processing is executed through quantum gates, depicted by unitary matrices $U$. 
These matrices are crucially unitary, meaning the complex conjugate transpose of $U$, denoted as $U^{\dagger}$, is also its inverse, fulfilling the condition $U^{\dagger}U=I$. 
This ensures the preservation of the inner product between vectors. 
However, extracting a conventional outcome from quantum processes necessitates measuring the qubits, which results in the collapse of their superposed state. 
In essence, practically speaking, QC involves manipulating the quantum state of qubits to, for example, increase the likelihood of obtaining a specific solution.

\subsection{Embedding classical data into quantum states}
In QC, a prevalent method for integrating classical data into quantum states is called basis embedding. 
This technique converts classical data, formatted as bit strings, into representations within quantum states.

\begin{definition}[Basis Embedding]\label{def:basisEmbed}
Given a sequence of bits $\vect{a} = a_1a_2...a_n$, basis embedding is the process of translating this sequence into the form $\ket{\vect a} = \ket{a_1} \otimes \ket{a_2} \otimes ... \otimes \ket{a_n}$, with $\ket{a_1a_2...a_n}$ serving as a more compact notation.
\end{definition}

Additionally, QC utilizes amplitude embedding, which encodes data into the amplitudes of quantum states, further expanding the range of possible computations. 

\begin{definition}[Amplitude Embedding]
    Given a normalized vector $\vect{a} = (a_1, a_2, ..., a_N)$ where $a_i \in \mathbb{R}$, $N = 2^n$ and $\sum_{i=1}^{N} |a_i|^2 = 1$, amplitude embedding is the process of mapping this vector into a quantum state $\ket{\vect{a}} = a_1\ket{0} + a_2\ket{1} + ... + a_N\ket{N-1}$, where each amplitude $a_i$ corresponds to the probability amplitude of the quantum state in the computational basis states $\ket{0}, \ket{1}, ..., \ket{N-1}$.
\end{definition}

Other embedding methods, such as angle and Hamiltonian embedding, are also employed in QC to represent data in various quantum-friendly formats.

\subsection{Certifiable robustness in machine learning}

Consider a binary classifier $f: \mathcal X \rightarrow \{0,1\}$ that assigns an input vector $\vect x \in \mathcal X$ (Euclidean space) to a binary class $f(\vect x) \in \{0,1\}$.
Additionally, we define as $y: \mathcal{X} \mapsto [0,1]$ its soft version which computes the model's logit value and the class prediction is made through $\mathbf{1} [y > 0.5]$.
Further, let \(o: \mathcal{X} \times \mathcal{X} \mapsto \{0,1\}\) denote the oracle that compares the semantic content of two input data points and returns $1$ if they are semantically the same. 
The research area of certifiable (provable) robustness aims to establish formal guarantees that under all admissible attacks within the budget, none will alter the model's prediction. 
Formally, certification at a point $\vect x$ involves evaluating if,  
\begin{equation}\label{eq:certified_robustness}
    f(\vect{\tilde{x}}) = f(\vect{x}) \text{ for all } \vect{\tilde{x}} \in \mathcal{X}\ \text{s.t.}\ o(\vect{\tilde{x}}, \vect{x}) = 1.  
\end{equation}

The assessment of semantic similarity between data points is complex and varies by task. 
This complexity is mitigated by considering two inputs semantically similar if their $\ell_p$-norm difference is less than a small threshold $\epsilon$, translating the robustness certification problem at $\mathbf{x}$ into:
\begin{equation}\label{eq:2:certified_robustness_norm_based}
    f(\vect {\tilde x}) = f(\vect x) \text{ for all } \vect{\tilde x} \in \mathcal X\ \text{s.t.}\ \lvert\lvert \vect{\tilde x} - \vect{x} \rvert\rvert_p \leq \epsilon.  
\end{equation}

For neural networks, robustness verification can be structured as a mixed-integer linear program (MILP) based on the network's weights~\cite{tjeng2018evaluating}. 
However, verifying neural network properties, especially with ReLU activations, is proven to be NP-complete~\cite{katz2017reluplex}. 
As a workaround, approximation methods are used, which might not certify all non-altering perturbations~\cite{gowal2019}. 
Randomized Smoothing (RS)~\cite{cohen2019certified, salman2019provably} has emerged as a notable method in the relaxed certification category, enabling robustness assessment independently of the classifier’s details.

\subsection{Randomized smoothing}
Randomized Smoothing  as proposed by \citet{cohen2019certified, salman2019provably}, constructs a smoothed classifier $g$ by evaluating the original classifier $f$'s logits $y$ around the input, defined as:
\begin{equation}\label{eq:rs_def}
    \begin{aligned}
        g(\vect x) = \mathbb E_{\vect z \sim \normal(0, \sigma^2 I)}[f(\vect{x} + \vect z)].
    \end{aligned}
\end{equation}

Here, the focus shifts from a direct prediction at $\vect{x}$ to the expected prediction over noisy samples distributed normally around $\mathbf{x}$. 
In practice, the goal is to determine the minimum value of $g$ over a ball of radius $\epsilon$ centered at zero in the $\ell_p$ space, and check if this minimum value exceeds \(\frac{1}{2}\), formally defined as:
\begin{equation}\label{eq:rs_problem}
     \min_{\bm{\delta} \in \mathcal{B}_p(\epsilon)} g(\vect{x} + \bm{\delta})\ \text{s.t.}\ \mathcal B_p(\epsilon): = \{\bm{\delta} \in \mathcal X : \lvert\lvert \bm{\delta} \rvert\rvert_p \leq \epsilon\}.
\end{equation}

Thus, let $\Phi$ denote the CDF of the standard normal distribution $\normal(0,1)$, and $\underline{p_1}$ be a lower bound of \cref{eq:rs_problem}, then $g(\vect x)$ as constructed in \cref{eq:rs_def} is certifiably robust if $\epsilon < \sigma\Phi^{-1}(\underline {p_1})$ for $1D$ classifier, i.e.,  $\mathcal X = \mathbb R$ \cite{cohen2019certified}.

\section{Robustness in QML}\label{sec:qml_robustness}

\begin{table*}[htb]
    \caption{List of encoding types, noise types and their respective input space interpretation. An indication of suggested pairings is given. Note that we do not further specify constraints (e.g., fixed Hamming weight distance for bitflip noise). We assume initialisation in the $\ket{000\dots}$-state.
    \label{tab:noise_vs_encoding}
    }
    \centering
    \begin{tabular}{llcc}
        \toprule
        {\bf Encoding} & {\bf Perturbation} & {\bf Input Space Interpretation} & {\bf Suggested} \\ 
        \midrule \midrule
        amplitude & depolarization noise & average feature contrast diminished & \checkmark \\ 
                  & amplitude damping & not intuitive &  \\ 
                  & gaussian noise & noisy features, $l_2$ adversarial defense & \checkmark \\ 
                  & uniform noise & noisy features, $l_1$ adversarial defense & \checkmark \\ 
                  & rotation noise & not intuitive &  \\ 
                  & bitflip noise & permutation of features &  \\ 
        \midrule
        rotation  & depolarization noise & feature contrast diminished and occurrence of previously not accessed states &  \\ 
                  & amplitude damping & pushes all features to $0$ &  \\ 
                  & gaussian noise & noisy features and occurrence of previously not accessed states &  \\ 
                  & uniform noise & noisy features and occurrence of previously not accessed states &  \\ 
                  & rotation noise & noisy features & \checkmark \\ 
                  & bitflip noise & not intuitive &  \\ 
        \midrule
        basis state & depolarization noise & all other data values gain small likelihood &  \\ 
                  & amplitude damping & not intuitive  &  \\
                  & gaussian noise & all other data values gain small likelihood stochastically & \\ 
                  & uniform noise & all other data values gain small likelihood stochastically &  \\ 
                  & rotation noise & not intuitive &  \\ 
                  & bitflip noise & randomly shift to other value, can be constrained to nearby values & \checkmark \\ 
        \bottomrule
    \end{tabular}
\end{table*}

Various approaches for certifiable robustness have been suggested for QML. In this extended section, we outline known methods for issuing robustness certificates and highlight their usage of perturbations. 
This leads to a conversation about the importance of jointly designing data encoding and perturbations.

\subsection{Existing work on robustness certificates in QML}

Weber~\textit{et al.}~\cite{weber2021} draw a theoretical correlation between binary quantum hypothesis testing (QHT) and certifiable robustness: while QHT aims at optimally distinguishing two quantum states, robust classification seeks to provide provable guarantees that two states are \textbf{not} distinguishable. Based on this connection, they derive a robustness bound for quantum classifiers subject to worst-case input noise, which they prove to be tight for binary classification. This bound, however, only holds in the regime where the probability of measuring the true class probability $p_A$ is always larger than the runner-up probability estimate $p_B$, which boils down to assuming perfect classification accuracy in an unperturbed setting. 
Such an assumption is common and necessary within the field of randomized smoothing to obtain robustness certifications, as seen in prior works~\cite{cohen2019certified, salman2019provably}.
However, this assumption may not be reasonable for machine learning models that are often overly confident and vulnerable to various uncertainties in practical applications. To address the limitations of assuming an ideal classifier, alternative approaches, e.g. conformal prediction~\cite{gendler2021adversarially, yan2023provably}, might be required.

\medskip

Another critical aspect we want to emphasize is the need of carefully considering the type of data encoding for the robustness bounds in \cite{weber2021} to provide meaningful insights on the trust that can be  given to the prediction of a perturbed data point. 
More precisely, as the bounds are based on the fidelity between input states, only non-zero fidelities allow useful conclusions on robustness. 
For example, when using basis state encoding, even the smallest changes in the input data result in zero fidelity, thus prohibiting any statement about predictions based on altered data.
While this aspect can be easily resolved by employing, e.g., the widely used rotation encoding, there is one more issue we want to highlight, which is related to the scalability of fidelity. 
As demonstrated by \citet{Zfiyczkowski2005}, fidelities between random states vanish exponentially in the number of qubits and hence even perturbations of small amplitude can cause a severe drop of certified robustness. 
This includes disturbances that lead to a low level of fidelity, yet may not directly impact the outcome of the algorithm we are analyzing.
The last point of criticism is best explained with a pathological example: Consider a binary classification algorithm $\mathcal{A}$ defined on a parameterized quantum circuit with two distinct subsystems, denoted $A$ and $B$. The algorithm maps inputs $x$ to label 1 if the probability of one of the qubits in subsystem $A$ being in the $\ket{1}$ state is larger than the corresponding probability for subsystem $B$:
\begin{equation}
  \mathcal{A}(x)=\begin{cases}
    1, & \text{if $\bra{\Psi} M_A^\dagger M_A \ket{\Psi} > \bra{\Psi} M_B^\dagger M_B \ket{\Psi}$}.\\
    0, & \text{otherwise}.
  \end{cases}
\end{equation}

For simplicity, consider three qubits with the first two constituting subsystem $A$ and the last one subsystem $B$. The measurement operators then take the form:
\begin{align*}
    M_A & = \bra{100} + \bra{110} + \bra{010} + \bra{000} \\
    M_B & = \bra{101} + \bra{111} + \bra{011} + \bra{001}
\end{align*}

Assuming the state $\ket{100}$ is to be certified, a perturbation could manipulate the state as far as $\ket{010}$, resulting in a fidelity of zero and thus a trivial certification radius of zero according to Weber~\textit{et al.}~\cite{weber2021}. At the same time, the algorithm would still be no less certain to return label 1 (the perturbed state is no closer to the decision boundary $\bra{\Psi} M_a^\dagger M_a \ket{\Psi} > \bra{\Psi} M_b^\dagger M_b \ket{\Psi}$).

This example is admittedly constructed, but consider generalized versions of it on larger systems. For instance, this might include a classification task over many input features, where only a few are indicative of the label to be returned, e.g., because only part of an image contains the object which we try to classify. Then there is lots of room for perturbations to push the state somewhere the classification outcome is unaffected, although fidelities might drop even exactly to zero. This impedes the utility of fidelity based robustness bounds.

In summary, the bounds derived in \cite{weber2021}, while valid, can be expected to be very conservative in practice, abstaining from certification in many cases where fidelities drop, simply because of the encoding used or the size of the system. This calls out for extensions which give tighter, more useful robustness statements in practice.

\medskip

A different line of research investigates the impact of various types of hardware noise on robustness bounds. 
Huang \textit{et al.}~\cite{huang2023} show that random rotation noise can improve robustness in quantum classifiers against adversarial attacks by establishing a connection to differential privacy. 
A certified robustness radius is derived in terms of the output probability ratio for binary classification given that the trace distance between perturbed and unperturbed state is smaller than some threshold. 
Using similar scaling arguments as before, we remark that the trace distance between two random quantum states vanishes exponentially in the system size and thus the condition for the robustness bound is violated easily for a large number of qubits.

In a similar work, it was shown for depolarization noise that robustness of quantum classifiers can be enhanced by increasing levels of noise~\cite{du2021}. 
Here as well we argue that the type of noise limits the data encodings which can be used meaningfully. 
While it has a reasonable interpretation for amplitude encoding, this type of noise has a questionable, highly non-trivial effect on basis state encoded data. We further discuss the co-design of perturbation and encoding in \cref{Co-Design of data encoding and perturbation type}.

\medskip

\citet{sahdev2023} approach quantum randomized smoothing from a slightly different angle. 
They introduce an algorithm based on Grover's search providing robustness bounds with provable quadratic sample advantage over classical methods. Our approach closely follows~\cite{sahdev2023} for several reasons. First, it is flexible in terms of the algorithm used and the type of noise.  Second, it clearly motivates the use of quantum computers due to the reduced number of samples necessary for a given robustness certificate. However, we do note some shortcomings as well. First, the Grover oracle necessitates a basis state encoding of the data, which limits the range of data encoding strategies available. Second, not all relevant perturbation types can be straightforwardly implemented. At last, the interplay of encoding and perturbation type is not considered.

\subsection{Co-Design of data encoding and perturbation type}
\label{Co-Design of data encoding and perturbation type}
In this article, we investigate QML models applied to classical data. As such, attacks and perturbations are applied in input space, and the perturbed $x$ is encoded into the quantum circuit. 
Quantum randomized smoothing introduces disturbances to the quantum circuit to achieve a more reliable prediction by analyzing the prediction statistics from multiple altered versions of the same data point.

If one is interested in a worst case robustness, as in adversarial defense, bounded perturbations are assumed and an unstructured perturbation is applied, e.g., Gaussian or uniform noise. 
If randomized smoothing is performed with other, potentially more structured perturbations, valid certificates can still be obtained, although certification radii can become overly conservative, as worst case robustness often already is. 
To avoid overly cautious certificates, a perturbation applied on the quantum circuit should be designed such that the perturbed state could be obtained by directly encoding a perturbed input data point, i.e., there should be an equivalent perturbation applied directly to the input data domain that produces the same state as when applying the perturbation on the quantum computer to the encoded state of the unperturbed data.

A second case is the certification of robustness to certain types of perturbations or noises. If these are resulting from the quantum hardware, perturbations which do not have an equivalent input domain perturbation may be useful, e.g., as they directly reflect the hardware noise. We put our focus on the more common case of certifying input space robustness. In this case, again, perturbations on the quantum circuit do need to have an input space interpretation.

\medskip

With this necessity in mind, consider Tab.~\ref{tab:noise_vs_encoding} where a selection of encoding and perturbation types is listed. For each combination, an interpretation of the induced changes on the state vector in input space is indicated. We leave it to the reader to work through all combinations and only give one example here: Consider amplitude encoded data acted upon by a depolarization channel. We can write the perturbed state as
\begin{align*}
    \ket{\tilde{\Psi}} & = (1-p) \sum_i^N a_i \ket{i} + \frac{p}{N} \mathbf{I} \\
                       & = \sum_i^N \left( (1-p) a_i + \frac{p}{N} \right) \ket{i} \\
                       & = \sum_i^N \tilde{a}_i \ket{i}
\end{align*}
from which we can see that the feature contrast, the difference between any two features $\tilde{a}_i$ and $\tilde{a}_j$ is diminished by a factor of $1-p$ compared to the contrast of unperturbed $a_i$ and $a_j$ and disappears completely as $p$ approaches $1$. Additionally, we see that $\ket{\tilde{\Psi}}$ can again be written as an amplitude encoded state. Hence, there exists an equivalent classical perturbation which could be used to obtain the $\tilde{a}_i$ coefficients.

As highlighted in the last column of Tab.~\ref{tab:noise_vs_encoding}, only few of the combinations of data encoding and perturbation type suffice the requirement of interpretability outlined above. Many others either access states which could not be reached by an equivalent classical perturbation with subsequent encoding, or lack a clear intuition of the induced changes. In quantum randomized smoothing, we suggest to be aware of the interplay of encoding and perturbation and design them simultaneously so that meaningful and not unnecessarily loose certificates are obtained. Furthermore, note that for a basis state encoding only bitflip noise allows a clear interpretation. We thus dedicate the next section to its implementation with Hamming weight constrains for later usage in QuAdRo.

\section{$k$-Hamming Distant States}\label{sec:hamming}

In this section, we describe how to construct a distribution around a binary input $\vect x\in \{0, 1\}^{n}$ of $k$-Hamming distant states.
This differs from the method of preparing Dicke states, where a Dicke state is an equal-weight superposition of all $n$-qubit states with Hamming weight $k$~\cite{bartschi2019deterministic}.
In our approach, we prepare a state $\psi: \{0, 1\}^{n} \to \mathcal{D}(\mathbb{H})$ that depends on $\vect x$.
\begin{definition}[$k$-Hamming distant states]\label{def:k_hamming_states}
    Given a binary input $\vect x\in \{0, 1\}^{n}$, we define a constrained $k$-Hamming distance set as: 
    \begin{equation}
        \mathcal{D}_{n, k}(\vect x) = \{\tilde{\vect x} \in \{0,1\}^n\, :\, \norm{\vect x - \tilde{\vect x}}_{0} = k\},    
    \end{equation}
    where the zero norm\footnote{the zero "norm" is not truly a norm.} corresponds to the non-zero elements differing from $\vect x$. 
\end{definition}
For any given positive integer Hamming weight $k$, with $k \leq n$, the number of states that are exactly $k$-Hamming distances away from a given state is given by the binomial coefficient $\binom{n}{k}$.
We can formalize the approach for allocating probabilities to states based on their Hamming distances from the input state, including a maximum probability for the input state itself ($i=0$) and decreasing probabilities for states with increasing Hamming distances ($i = 1$ to $k$).
To do this, we need to ensure that:
\begin{enumerate*}[label=(\roman*)]
    \item the input state ($i=0$) has the highest probability,
    \item probabilities for states decrease as their Hamming distance from the input state increases and 
    \item probabilities are equally distributed among states at the same Hamming distance.
\end{enumerate*}.
First, we define a base weight of $e^{-\frac{i}{\sigma}}$ for each Hamming distance $0 \leq i \leq k$, including a special consideration for $i=0$ to ensure it receives the maximum weight of 1. 
Here, $\sigma > 0$ works as hyper-parameter that adjusts the distribution of weights.
A larger $\sigma$ would result in a slower decrease in weights with increasing $i$, allowing for more uniform distribution of probabilities across different Hamming distances.
Thus, the normalized weight for each Hamming distance $i$ is given by:
\begin{equation}
    w(i) = \frac{e^{-\frac{i}{\sigma}}}{\sum_{i=0}^{k} e^{-\frac{i}{\sigma}}}\ \text{for all}\ 0 \leq i \leq k. 
\end{equation}

Finally, to assign probabilities to individual states at a given Hamming distance $i$, each state receives a fraction of the total probability allocated to that distance:
\begin{equation}
    p(i) = \frac{w(i)}{\binom{n}{i}} \quad \text{for each state at distance } i.  
\end{equation}

As an example, let us consider an input vector $\vect x = (0, 1, 1)$ which has a length of $n=3$; there are exactly three states that are Hamming distance of 1 away from it, three of distance 2 and one state of distance 3.
For $\sigma=1$, the distribution of probabilities for states based on their Hamming distance is: 0) the initial state with probability of $0.644$, 
\begin{enumerate*}
    \item 3 states, each with a probability of $0.079$,
    \item 3 states, each with a probability of $0.029$ and 
    \item 1 state with a probability of $0.032$.
\end{enumerate*}

\subsection{Construction of $k$-Hamming distant states}\label{sec:hamming_construction}

Given an input $\vect x\in\{0, 1\}^n$, we construct a superposition of $k$-Hamming distant states from $\vect x$ in the Hilbert space as:
\begin{equation}\label{eq:hamming-state}
    \ket{\psi(\vect x)} = \sum_{i = 0}^{k} \sqrt{p(i)} \sum_{\tilde{\vect x} \in \mathcal{D}_{n, i}(\vect x)} \ket{\tilde{\vect x}}.
\end{equation}
These can be understood as the probabilities of staying in the original state $\vect x$ or transitioning to the nearest state that is $k$-Hamming distant away.

A direct approach to construct the quantum state $\ket{\psi(\vect x)}$ involves first calculating the probabilities associated with the $k$-Hamming states. 
Following this step, the quantum circuit is prepared using the M\"ott\"onen state preparation method~\cite{mottonen2005transformation}, a widely recognized technique for incorporating data into the amplitudes of quantum states. 
Despite its effectiveness, this approach encounters a significant limitation due to the exponential increase in the depth of the circuit, symbolized by $\mathcal{O}(2^n)$, which pertains to the exponential growth in the requirement for both CNOT and single-qubit rotational gates as the number of qubits increases. 
This exponential increase presents an additional challenge, specifically when attempting to enumerate classically all the probabilities corresponding to states that are $k$-Hamming distances apart. 
As the dimensionality of the input grows, this task becomes increasingly unmanageable, highlighting a critical bottleneck in the scalability of this method for larger quantum systems.

\begin{figure}
    \centering
    \subfigure[Circuit preparation with controlled RY($\sigma\pi$).]{
        \includegraphics[width=0.45\textwidth]{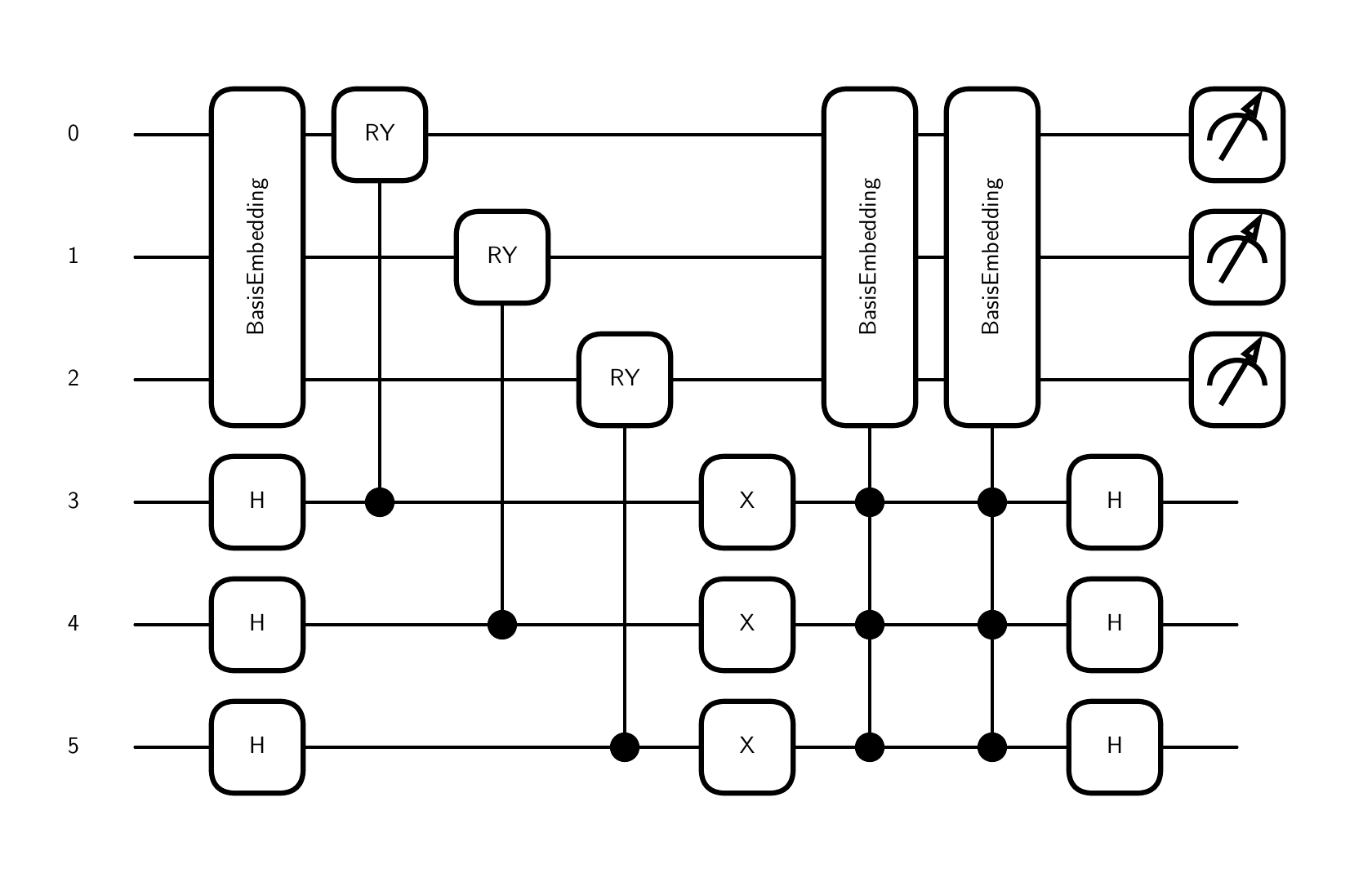}}
    \subfigure[Quasi-probability distributions for varying $\sigma$.]{
        \includegraphics[width=0.45\textwidth]{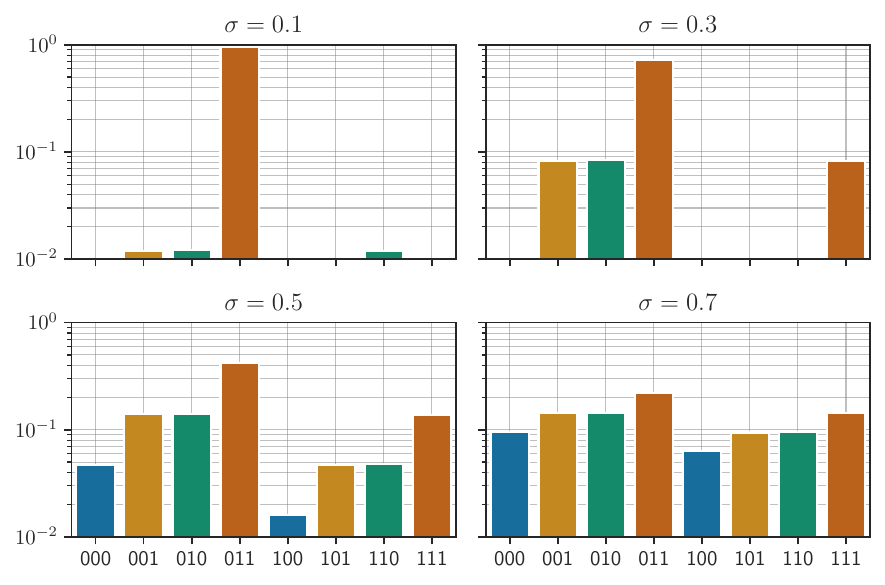}}
    \caption{1-Hamming distant state preparation for $\vect x=(0, 1, 1)$.}
    \label{fig:hamming-state-preparation}
\end{figure}

\medskip
\paragraph{Our state preparation}
To construct the quantum state $\ket{\psi(\vect x)}$ effectively, we propose a method that includes the use of extra qubits, termed ancillary qubits, along with controlled rotational gates. 
This method is enhanced by introducing a technique that involves the inverse application of multi-controlled basis embedding for the input. 
This enables the design of a more efficient circuit that requires the addition of only one qubit for every input dimension. 
In \cref{fig:hamming-state-preparation}, we showcase the procedure for setting up a circuit for a three-dimensional input. 
The process begins with the initial preparation of the $x$ state using a basis embedding technique. 
Following this, $n$ ancillary qubits are introduced, and a combination of Hadamard gates and controlled RY gates, adjusted by $\sigma \pi$, are applied.
In this scenario, $\sigma \in [0, 1]$ and an increase of its value denotes an overall increase on the likelihood of neighbouring states.
We observe that for $\sigma = 0.5$ the controlled RY rotation corresponds to a controlled Hadamard, hinting the quantum Fourier transform state preparation.
Finally, the ancillary qubits are manipulated with a series of PauliX gates before the input undergoes a dual embedding in the basis via a multi-controlled gate.
Inverting the ancillary qubits prior to introducing a controlled input decreases the likelihood of states that are distant from the input.

\medskip
Unlike the M\"ott\"onen state preparation method, our technique does not replicate the distribution with exact precision. 
In practical terms, this situation means that while our method can effectively adjust the likelihood of certain quantum states being realized - specifically, those within a $k$-Hamming distance from a given state - it also unintentionally boosts the chances of encountering states beyond this specified Hamming distance.
Essentially, when we attempt to increase the probability of closely related states (those a few quantum flips away), we inadvertently also make it more likely to observe states that are further away than intended. 
This can introduce inaccuracies in quantum simulations or computations that rely on precise probability distributions.

\begin{figure}
    \centering
    \subfigure[Circuit preparation for a uniform distribution.]{
        \includegraphics[width=0.45\textwidth]{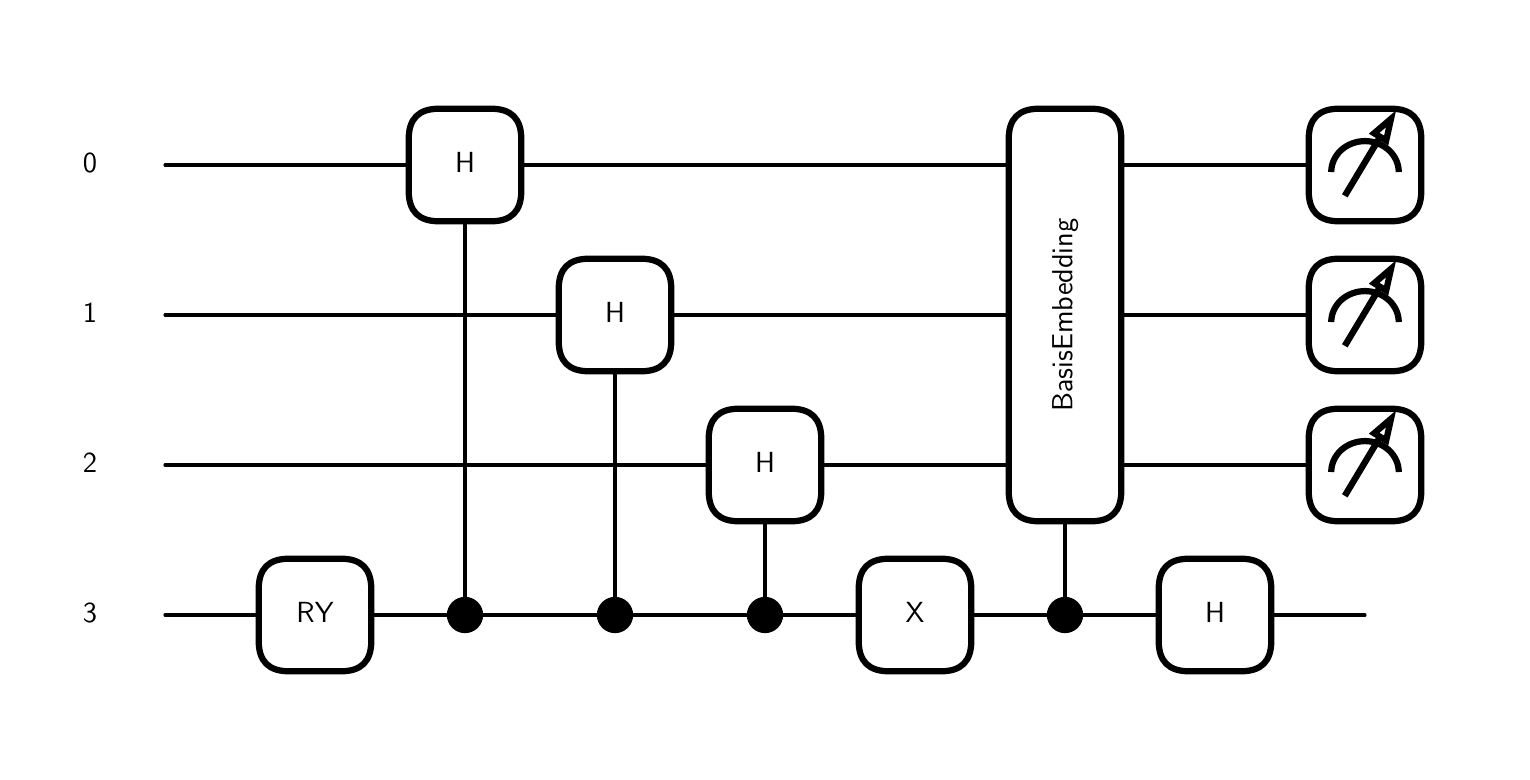}}
    \subfigure[Quasi-probability distributions for varying $\sigma$.]{
        \includegraphics[width=0.45\textwidth]{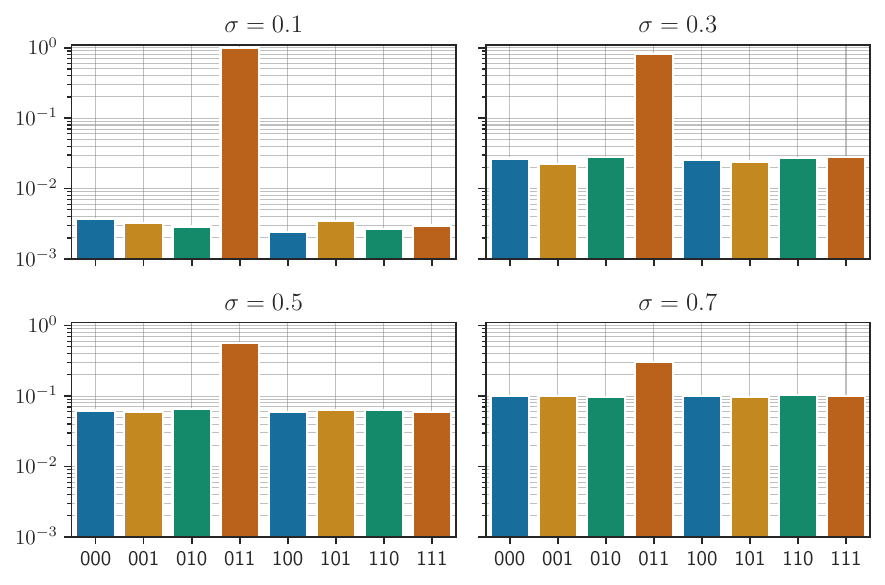}}
    \caption{Uniform distant state preparation for $\vect x=(0, 1, 1)$.}
    \label{fig:uniform-state-preparation}
\end{figure}

\medskip
\paragraph{Uniform state preparation}
In \cref{fig:uniform-state-preparation}, we detail the process for setting up a uniform distribution centered on a three-dimensional input along with its corresponding probabilities. 
Continuing from the earlier discussion, the probabilities for adjacent states are uniformly likely across all states. 
This method of state preparation is more efficient since it only necessitates a single ancillary qubit, and we can also adjust $\sigma$ to boost the probability of nearby qubits through a RY gate on the ancillary. 
Nonetheless, treating the probability of neighboring states as equally likely for all states presents a drawback when considering semantically meaningful perturbations for the input state.

\subsection{Certified robustness for $k$-Hamming distant states}\label{sec:certified_robustness}

Here, we formally describe the computation of the certified robustness distance in relation to a distribution that is $k$-Hamming distant from the given input.
We focus on the robustness certificate based on the $\ell_0$-norm within a binary input domain $\mathcal{X} = \{0, 1\}^n$. 
Specifically, let $p_i:\mathcal{X}\to [0, 1]$ denote the likelihood that smoothing the base classifier $f$ with a $k$-Hamming distant distribution around $\vect x$ yields class $i$, formally:
\begin{equation}\label{eq:hamming_smooth_classifier}
    p_i(\vect x) = \E_{\tilde{\vect x} \in \mathcal{D}_{n, k}(\vect x)} [f(\tilde{\vect x}) = i],
\end{equation}
where $\mathcal{D}_{n,k}(\vect x)$ represents the set of $k$-Hamming distant vectors for $\vect x$.
As established by \citet{lee2019tight}, for any input $\tilde{\vect x}$ with $\norm{\vect x - \tilde{\vect x}}_0 \leq k$ across all classes $i\in \mathcal{K}$, the following holds:  
\begin{equation}\label{eq:certified_robustness_distance}
    |p_i(\tilde{\vect x}) - p_i(\vect x)| \leq \Delta,\; \text{where}\; \Delta = \argmax_r p_r \geq 0.5,
\end{equation}
defines the certified robustness distance.
Based on the architecture of the base classifier, directly computing $p_i(\vect x)$ might not be feasible.
However, we can instead generate a finite set of samples from $\mathcal{D}_{n,k}$, in order to estimate a lower bound $\underline{p_i(\vect x)}$ with high confidence ($1-\alpha$).
In practice, for an input $\tilde{\vect x}$ with $\norm{\vect x - \tilde{\vect x}}_0 \leq k$,
\begin{equation*}
   \text{if}\quad \underline{p_i(x)} - \Delta > \frac{1}{2},\quad \text{then}\quad \argmax_j p(x)_j = i,
\end{equation*}
with probability at least $1-\alpha$~\cite{cohen2019certified}.

\medskip
In practice, to establish a lower bound for $\underline{p_i(\vect x)}$, traditional methods employ the Clopper-Pearson interval~\cite{cohen2019certified}.
Nonetheless, the quantity of samples, which corresponds to the number of requests made to the base classifier, increases following an O(1/$\alpha^2$) complexity, where $\alpha$ represents the target error margin within the expected probability value, delineated over the randomized smoothing area surrounding the input. 
In the context of QML, \citet{sahdev2023} demonstrate that for achieving the same level of confidence and error margin, the requisite number of inquiries to the base classifier is reduced to $\mathcal{O}(1/\alpha)$.
QuAdRo~\cite{sahdev2023} employing $m$ ancilla qubits for the quantum phase estimation allows for a maximum of $M = 2^{m+1}-1$ oracle calls.
In addition, the likelihood of success approaches nearly 100\% by conducting multiple trials and selecting the median outcome.
\section{Experimental Results}\label{sec:experiments}

\subsection{Setup}\label{sec:setup}

\begin{figure}
    \centering
    \includegraphics[width=0.45\textwidth]{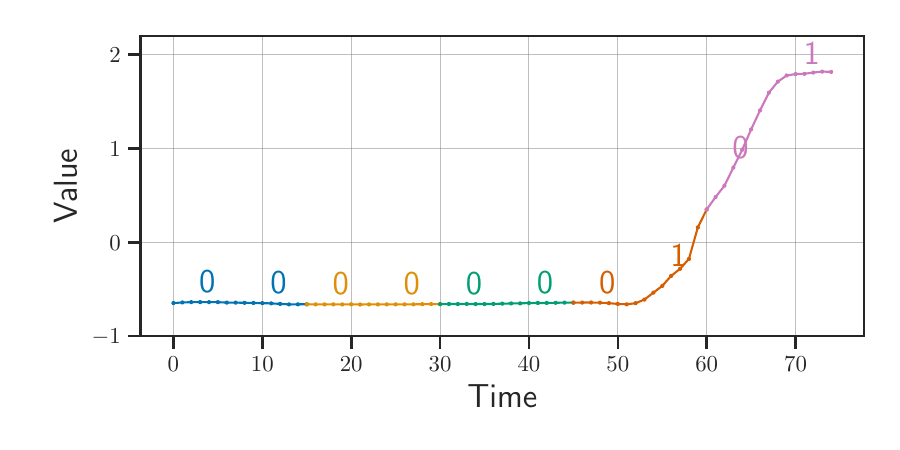}
    \vspace{-1em}
    \caption{Binary representation of discretized patterns in time series, visualized through the Bag-of-Words representation. The method operates by splitting in half a sample from the gun-point~\cite{UCRArchive} time series dataset, then dividing into windows of size 15 and bins of two with a word size of 2 (binary).}
    \label{fig:bow_representation}
\end{figure}

\paragraph{Datasets}
In this experiment, we consider two datasets: 
\begin{enumerate*}[label=(\roman*)]
    \item \texttt{Iris}~\cite{iris} and
    \item \texttt{GunPoint} from the UCR time series~\cite{UCRArchive} archive
\end{enumerate*}.
In preprocessing the \texttt{Iris} dataset, the \textit{versicolor} class is removed, \textit{virginica} is mapped to 0 and \textit{setosa} to 1. 
The input feature \textit{petal width} is excluded, all samples are normalized to unit norm, and then split into a 60/40 training/testing ratio.
In the preprocessing of the \texttt{GunPoint} dataset from the UCR time series archive~\cite{UCRArchive}, we consider using a Bag-of-Words~\cite{weinberger2009feature, lin2012rotation, JMLR:v21:19-763} approach to create a new representation of the data with reduced dimensionality.
Initially, the dataset is divided into training and testing sets, from which only the first half of each time series is selected for further processing. 
This truncation serves to focus on a specific portion of the time series that is of interest.
The Bag-of-Words model is then applied to these truncated time series as shown in \cref{fig:bow_representation}. 
The model operates by dividing each time series into windows of a specified size (15 in this case) and encoding the data within each window into \textit{words} based on a predefined number of bins (2 bins here) and a word size of 2.
The resulting data is a binary vector of size 10, i.e. $x \in \{0, 1\}^{10}$.
This representation of time series can also be made more expressive by increasing the number of bins and the word size. 
Since it remains quantized by design, a mapping to a basis state encoding can always be undertaken easily.
\medskip

With the Bag-of-Words preprocessing, input space perturbations appear as bit-flips of the binary data representation. 
Specifically, if a region of the data snippet is perturbed sufficiently strong that its amplitudes falls into a different bin, the corresponding bit is flipped. 
By constraining the number of bits that are flipped, i.e., the Hamming weight distance, the magnitude of the amplitude deviations induced by the perturbation can be controlled. 
Thus, we conclude that Hamming weight constrained bit-flip noise is particularly suitable for this type of basis state encoded data.

\medskip
\paragraph{Network Architecture}

In our experiments, we employ a QNN to function as an oracle in Grover's algorithm. 
This is achieved by incorporating a \textit{strongly entangling layer}~\cite{schuld2020circuit} as the primary architecture component. 
Given that our datasets are simplified to binary classification problems, we have trained the network on a singular output qubit to perform the classification.
In \cref{tab:network_architectures}, we report the accuracy, number of qubits and layers for each dataset considered.

\begin{table}[htb]
    \centering
    \caption{QNN architecture's parameters.}
    \label{tab:network_architectures}
    \begin{tabular}{lllll}
    \toprule
    Dataset     &Accuracy   &Embedding  &N. of Qubits &N. of Layers \\
    \midrule
    Iris        &1.00       &Basis      &3          &2 \\
    Gun Point   &0.92       &Basis      &10         &50  \\
    \bottomrule
    \end{tabular}    
\end{table}

\medskip
\paragraph{QuAdRo Parameters}

In our application of QuAdRo~\cite{sahdev2023}, we employed the \textit{iterative quantum phase estimation}~\cite{iqpe} technique to decrease the number of ancillary qubits needed for phase estimation. 
Given $M$ as maximum number of calls to the oracle, we consider $m = \ceil{\log_2(M)}$.
As a result, we execute the circuit $m$ times, with each run involving $\sqrt{N}$ shots, in contrast to the $N$-times required by the standard implementation of randomized smoothing.
For each iteration $i$, we evaluated the $2^i$ power of Grover's algorithm and estimated the phase using a classical method.
Employing all available $m$ qubits for phase estimation allows for executing the circuit with only $\sqrt{N}$ repetitions.
For determining the minimum value of $p(x)$, we adopt a 90\% confidence level ($\alpha = 0.1$) using the Clopper-Pearson interval.

\begin{figure}
    \centering
    \includegraphics[width=0.4\textwidth]{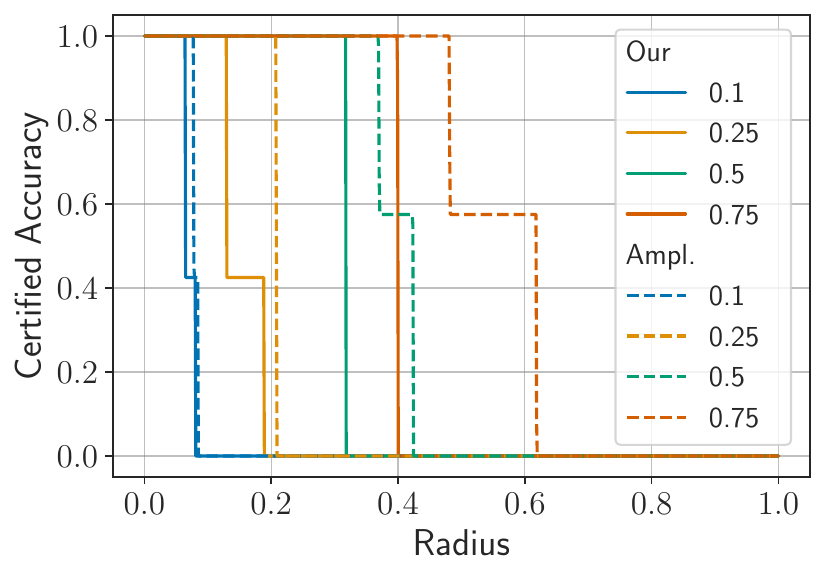}
    \caption{Certified accuracy of quantum randomized smoothing with perturbations limited to a Hamming distance of 1 from the original input, across different $\sigma$ values. Each evaluation maintained a consistent number of 10$^9$ shots.}
    \label{fig:certified_for_varying_sigma}
\end{figure}

\subsection{Evaluation of 1-Hamming distant state distributions}

To conduct a practical comparison of the probability outcomes derived from various methods of creating states with $k$-Hamming distance, we opt to measure the certified accuracy of a QNN that has been trained on the Iris dataset. 
In \cref{fig:certified_for_varying_sigma}, we showcase an assessment of quantum randomized smoothing, utilizing a fixed number of trials (one billion shots), to examine how certified accuracy varies with different values of $\sigma$ while maintaining a 1-Hamming distance. 
This analysis highlights the impact of varying the probability of a state flipping to another state that differs by a Hamming distance of 1.
Generally, higher values of $\sigma$ indicate a greater tolerance to state changes, reflected in the certified accuracy across the radius spectrum. 
While differences between amplitude encoding and our method of state preparation are noticeable at larger radii, both methods generally demonstrate similar behavior, emphasizing the commonalities in how states are prepared.

\begin{table}[!htb]
    \centering
    \caption{Circuit preparation for 1-Hamming distant states.}
    \vspace{-0.5em}
    \label{tab:depth_hamming}
    \begin{tabular}{llll}
        \toprule
        State           &N. of qubits   &N. of gates &Depth  \\
        \midrule
        Ampl. (M\"ott\"onen~\cite{mottonen2005transformation})    &10             &2045        &2036 \\
        Our             &20             &43          &6 \\
        \bottomrule
    \end{tabular}
\end{table}

As indicated in \cref{tab:depth_hamming}, there is a considerable difference in circuit depths between the two methods: the  M\"ott\"onen amplitude state preparation technique results in a circuit depth of 2036, whereas our method for state preparation yields a much shallower depth of just 6 when applied to the GunPoint dataset, which features binary input with a dimensionality of 10.
Despite this, the quantum resource requirement in terms of qubits for our approach is twice as high.
In general, our method requires only $\mathcal{O}(2n)$ gates for $n$ qubits, while M\"ott\"onen's state preparation scales exponentially at $\mathcal{O}(2^n)$.

\subsection{Comparison between uniform and 1-Hamming distant state distributions.}

\begin{figure}
    \centering
    \subfigure[1-Hamming distribution.]{
        \includegraphics[width=0.4\textwidth]{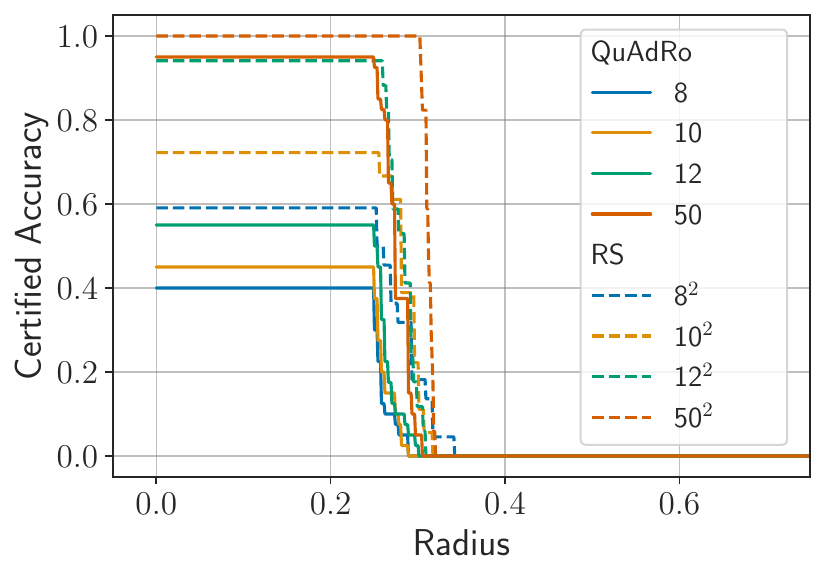}
    }
    \subfigure[Uniform distribution.]{
        \includegraphics[width=0.4\textwidth]{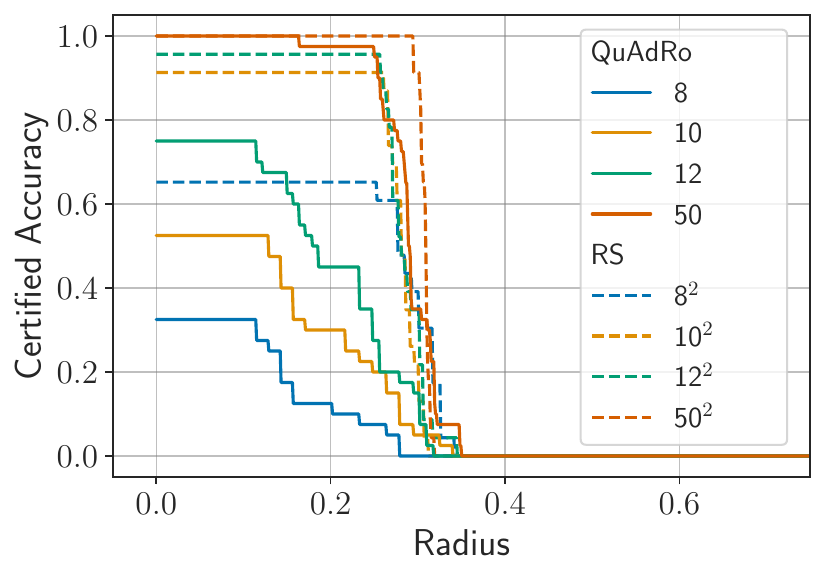}
    }
    \caption{Certified accuracy of quantum randomized smoothing between QuAdRo and RS in terms of shots number for 1-Hamming distant and uniform distributions. Both tests have been conducted with a $\sigma$ of 0.5 for the Iris dataset.}
    \label{fig:iris_for_varying_shots}
\end{figure}

Here, we compare the 1-Hamming distance distribution with respect to the uniform distribution in terms of certified accuracy.
For both scenarios, we execute a plain version of \textit{randomized smoothing} (RS)~\cite{cohen2019certified}, which involves a QNN placed ahead of the input distribution, running for a number of shots that is quadratically higher in comparison to QuAdRo.
Specifically, we run a set of shots $\{8^2, 10^2, 12^2, 50^2\}$ for RS and $\{8, 10, 12, 50\}$ for QuAdRo.
In the context of QuAdRo, we implement the iterative quantum phase estimation which makes use of one single additional qubit.
Consequently, the total quantum circuit consists of 6 qubits for the uniform distribution and 9 qubits for the 1-Hamming distribution.
We conduct our experiments in the \textit{default.qubit} simulator in PennyLane\footnote{PennyLane version 0.33.1 \href{https://github.com/PennyLaneAI/pennylane}{https://github.com/PennyLaneAI/pennylane}.}. 
\medskip

In \cref{fig:iris_for_varying_shots}, we show a comparison for $\sigma = 0.5$ for the Iris dataset. 
The analysis shows that the 1-Hamming distribution has a tendency to reach a higher level of certified accuracy when compared with the uniform distribution. 
This result is attributed to the increased probability of locating states in proximity to the initial state, as opposed to states dispersed uniformly at a distance.
In particular, for QuAdRo, achieving high certified accuracy requires fewer than 50 shots of the circuit, compared to 144 shots for RS. 
With enough shots, both distributions achieve a certified accuracy of 1. 
This outcome is because the Iris classifier achieves perfect clean accuracy and with $\sigma=0.5$, as demonstrated in \cref{fig:uniform-state-preparation} and \cref{fig:hamming-state-preparation}, maintains the probability of the original input higher than that of the neighbors.

\subsection{Certified robustness for time-series analysis}

\begin{figure}
    \centering
     \subfigure[1-Hamming distribution.]{
        \includegraphics[width=0.4\textwidth]{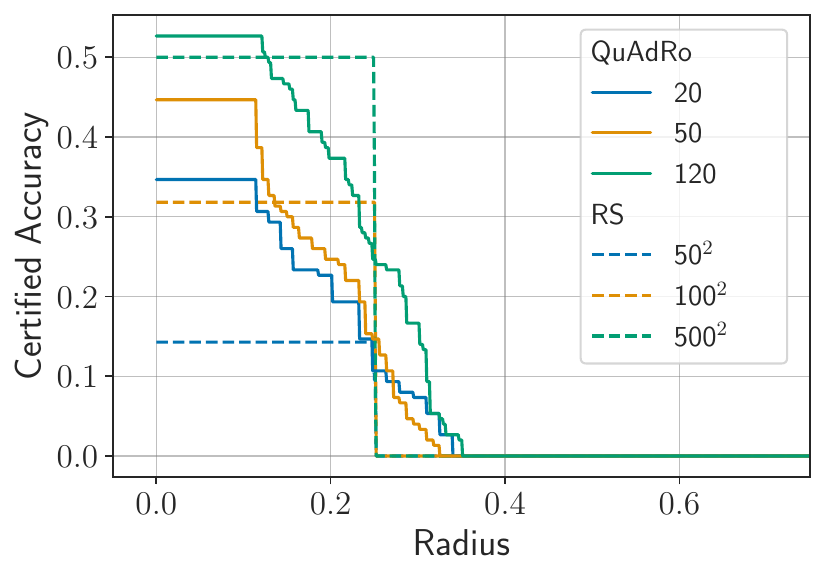}
    }
    \subfigure[Uniform distribution.]{
        \includegraphics[width=0.4\textwidth]{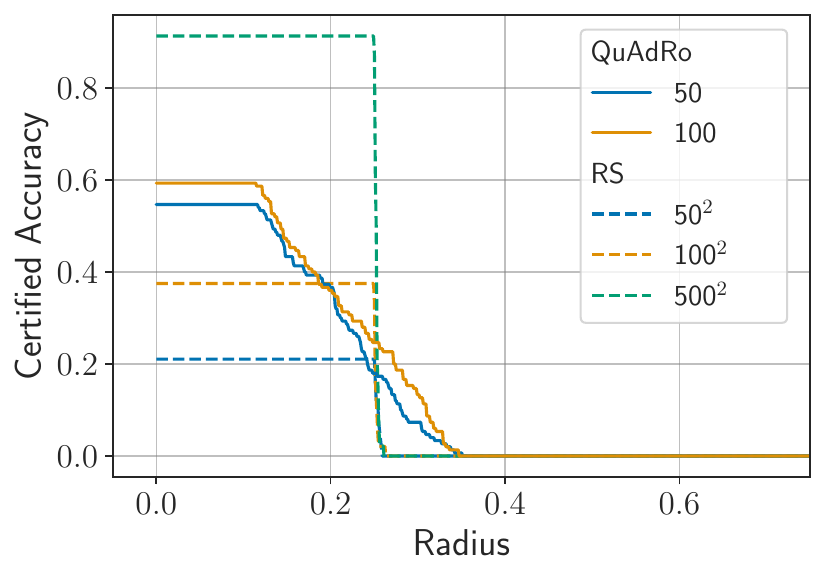}
    }
    \caption{Certified accuracy of quantum randomized smoothing between QuAdRo and RS in terms of number of shots for 1-Hamming distant and uniform distributions. Both tests have been conducted with a $\sigma$ of 0.5 for the GunPoint dataset.}
    \label{fig:gun_point_varying_shots}
\end{figure}

In \cref{fig:gun_point_varying_shots}, we present a comparison based on the GunPoint dataset with $\sigma = 0.5$.
As previously described, the input was transformed using a Bag-of-Words model, resulting in a binary format to encode the state within the basis. 
In addition, we consider the iterative quantum phase estimation, which requires 13 qubits for a uniform distribution and 22 qubits for a 1-Hamming distribution.
In the 1-Hamming distribution case (\cref{fig:gun_point_varying_shots}.a), QuAdRo consistently outperforms RS across all radii, with iterations of 20, 50, and 120 for QuAdRo compared against 50, 100, and 500 samples for RS.
Within the uniform distribution, we observe (\cref{fig:gun_point_varying_shots}.b) that QuAdRo demonstrates a quadratic improvement when executed for 50 and 100 iterations compared to RS.
\medskip

Due to the limited number of measurements, the phase estimation of QuAdRo does not precisely fit the exact radius. 
This causes QuAdRo to exhibit smoother performance compared to RS. 
The likelihood of accurately measuring the phase up to $m$ bits is $\frac{8}{\pi^2}$. 
By conducting the experiment repeatedly and applying the median estimate~\cite{miyamoto2022bermudan}, we can rapidly achieve a 100\% accuracy rate.
In practice, QuAdRo obtains probabilistic robustness certificates, similar to RS, but with a quadratic speed-up in the number of shots required.
The average run-time for the simulated circuit with a 1-Hamming distribution is roughly 21 minutes per sample on a server with a 128 CPU cores and 775 GB of RAM memory.

\section{Discussion of Experimental Results}\label{sec:discussion}

The challenges of implementing the quantum phase estimation algorithm are significant for NISQ devices.
To address these challenges, our approach includes several key considerations.
First, using the Bag-of-Words model to represent data helps reduce the dimensionality of time-series data while retaining crucial information for accurate classification.
Second, our construction of the $k$-Hamming distance for state preparation enables a relatively shallow circuit. 
Although this circuit approximates the exact distribution, the depth is three orders of magnitude smaller than that required for amplitude state preparation.
This reduction is crucial for the overall preparation process of the quantum phase estimation. 
Third, the use of iterative quantum phase estimation decreases the number of qubits needed for precise phase estimation.
This makes the QuAdRo algorithm more compatible with current quantum hardware, which is restricted by qubit stability and coherence times.

\medskip

Experimental findings demonstrate the efficacy of the 1-Hamming distance approach in achieving higher certified accuracy, especially in scenarios with limited shots.
This result highlights the advantages of using constrained perturbation models like the 1-Hamming distance in situations where it is reasonable to consider bit-flips as principal source of noise.
However, it is crucial to recognize that other sources of noise can significantly impact the practical implementation and benefits of robustness certification.
Gate noise and decoherence, in particular, pose additional challenges to the overall certification process, potentially affecting the algorithm's performance on real quantum hardware.
To address these broader noise concerns and enhance the QuAdro algorithm's performance on NISQ devices, a comprehensive strategy is essential.
Error mitigation techniques like zero-noise extrapolation can be applied to reduce noise effects on circuit outputs. Simultaneously, minimizing two-qubit gates through circuit layout optimization can enhance quantum operation fidelity. This combined approach aims to boost the algorithm's resilience to NISQ device noise, potentially improving its real-world performance and accuracy.
\section{Conclusion}

In this study, we establish a connection between input transformation and data encoding. 
Our analysis indicates overlooked opportunities in encoding techniques and the related noise types. 
Expanding on previous methods of preparing input distributions, we propose a new encoding strategy that generates $k$-Hamming distant states using fewer gates than needed for amplitude embeddings. 
Our method, however, does not exactly replicate the distribution of $k$-Hamming distant states and requires $n$ additional ancillary qubits, where $n$ is the number of qubits used for embedding the state. 
Despite this drawback, the complexity of our circuit is significantly reduced, being three orders of magnitude less deep than that required for amplitude state preparation. 
In our empirical validation, preparing $k$-Hamming states instead of uniform states resulted in higher certified accuracy for the same number of shots. 
In addition, using iterative quantum phase estimation demonstrates a quadratic improvement in speed compared to standard sampling in verifying the robustness of QNN.

\printbibliography

\end{document}